\documentclass[a4paper, 11pt]{article}
\RequirePackage[margin=15mm]{geometry}
\RequirePackage{fancyhdr}

\newcommand{\keywords}[1]{
\medskip
Keywords: \textit{#1}
}

\newcommand{\dedication}[1]{
\medskip
\textit{#1}
}

\newenvironment{affiliations}{
\medskip
}

\renewenvironment{abstract}{
\small
\medskip\medskip
}

\usepackage{datetime}
\usepackage[utf8]{inputenc}
\usepackage[noenc]{tipa}
\usepackage{amssymb,amsfonts,textcomp,amsmath}
\usepackage[T3,T1]{fontenc} 
\usepackage{csquotes}
\usepackage{hyperref} 
\usepackage{url}
\usepackage{amsmath}
\usepackage[english]{babel}
\hypersetup{colorlinks=true, linkcolor=red, citecolor=red, filecolor=blue,
  urlcolor=blue, pdftitle=, pdfauthor=Georges Sitja, pdfsubject=, pdfkeywords=}

\begin{document}

\pagestyle{fancy}
\fancyhead{} 
\fancyfoot{} 
\fancyfoot[R]{G.S. ~-~ \the\day-\the\month-\the\year ~-~\xxivtime}
\fancyhead[R]{\thepage}
\fancyhead[L]{\leftmark \\ \rightmark} 
\footskip = 40pt
\textheight = 730pt
\title{Statistics modification under monomer diffusion}

\date{\vspace{-5ex}} 

\maketitle


\dedication{}

\begin{affiliations}
G. Sitja\\
CNRS, UMR 7325, Aix-Marseille University, Cinam, Campus Luminy, Case 913,\\
F-13288 Marseille 09, France.
\end{affiliations}


\color{black}

\begin{abstract}
  The diffusion and coalescence of individual atoms on a nanostructured surface are treated in
  a purely statistical way. From this, analytical formulas are derived which, from a known
  initial state, give the final cluster size distribution on a surface after the diffusion
  of all the individual atoms. Unexpectedly, it turns out that these formulas allow obtaining
  a statistical law giving the size histogram of the clusters when only homogeneous nucleation
  occurs with a critical germ equal to 2, and in the situation where nucleation starts once
  the deposition of the atoms is completed.
\end{abstract}

\section{Introduction}
A growing interest in surface science is the study of the properties (optical, chemical...)
of metal clusters. These clusters can be synthesized by different methods. First synthesis
were done by  growing metal particles by condensation of metal vapor under UHV (Ultra High Vacuum)
on a metal-oxide single crystal or on a thin film on a metal single crystal
\cite{ISI:A1984ABP3500010}\cite{ISI:000075637300001}\cite{ISI:000180989100014}%
\cite{ISI:000175303400013}, but the size dispersion of such preparation were too large to study
size effects. In order to correct this imperfection new ways to synthesize metal-clusters,
such as the deposition of mass-selected particles \cite{ISI:000167650100015}%
\cite{ISI:000558827500023}\cite{ISI:000396146900021}, lithography \cite{ISI:000077542300197},
or the use of self-assembled colloidal masks \cite{ISI:000247076900007}, are used. For about
twenty years, researchers use new supports allowing the organization of particles in a regular
array during growth \cite{ISI:000188238900023}\cite{ISI:000440505000051}\cite{ISI:000350677000001}%
\cite{ISI:000318892400017} providing model systems with a much higher density than by size
selected clusters deposition and however with a sharp size distribution. These arrays can
be precisely characterized by surface science techniques like STM (Scanning Tunneling Microscope)
\cite{ISI:000488835000007}\cite{ISI:000458704800047}\cite{ISI:000440505000051}%
\cite{ISI:000318892400017}, GISAXS (Grazing-Incidence Small-angle Scattering)
\cite{ISI:000490353900017}\cite{ISI:000318892400017}.
The high density of clusters ($10^{12} - 10^{13}$ clusters per $cm^2$) reachable by these
preparation methods allows studying the physical and the chemical properties of few atoms
clusters down to the limit of individual atoms \cite{ISI:000444205500004}%
\cite{ISI:000318892400017}\cite{ISI:000458704800047}. The sharp size distribution of
particles is generally limited by the Poisson distribution \cite{ISI:000488835000007}%
\cite{ISI:000458704800047}\cite{ISI:000274979000024}. Moreover, few years ago, the world
of heterogeneous catalysis became interested by catalysis by individual atoms (the
so-called Single-Atom Catalysis - SAC) \cite{ISI:000292999100017}\cite{ISI:000433404300004}.
One can legitimately wonder what happens to the initial distribution if during the experiments
the individual atoms diffuse, as some authors are suggesting \cite{ISI:000471212600107}%
\cite{ISI:000600017400006}, by temperature increase or by interaction with reactants gases.
Experimental studies need perfectly well-characterized model systems, not only after
synthesis but also during reaction where coalescence could occur.

\section{Motivations}
This study is motivated by the following practical problem (which will serve as an
experimental reference when it is not specified otherwise): The deposition of metal
atoms to study their catalytic activity (for example, by condensing a flux of atoms
coming from an evaporator) on a surface exhibiting nucleation centers distributed on
an array. In this case, because each nucleation site is equivalent, the probability,
for a diffusing atom on the surface, to be captured is the same for all the nucleation
centers considered.  One of the first question that can be asked is : ``What is the size
distribution of the particles formed as a function of the quantity of atoms deposited?''.
In fact, the answer to this question is already known: After the deposition of an average
of x atoms per nucleation site, the probability of having a cluster of n atoms follows
Poisson's law \cite{Poisson}:$P(x,n)=\frac{x^n}{n!}e^{-x}$ . If the particles are stable,
there is no reason for these probabilities to change, however, as it is almost always the case,
the deposition conditions may be different from the conditions under which the properties of
these clusters are studied: The temperature may change, the chemical environment (the gases
used for the study) can vary, radiation (laser, UV source...) may be used... all these
changes can "destabilize" the arrangement of the particles by initiating diffusion. This
is the case for some works \cite{ISI:000458704800047}\cite{ISI:000490353900017}, and in particular
for the experiments described by
Düll \cite{ISI:000488835000007} for which I will make a small analysis, in the light of the 
formulas determined
in this article, in the section \ref{confrontation}. Generally speaking, the smaller
the size of a metal cluster, the easier it will diffuse (even if in some cases this may not
be true). One can expect that the clusters formed by a single atom (the monomers) will be
the first to start moving when environmental conditions change. Of course, if conditions
become so extreme that all metal particles either diffuse or evaporate, the experiments
loose their interest. The purpose of this paper is to give the probability law for the
cluster size distribution in the case where only single atoms can diffuse.

\section{Assumptions of this study}
\label{hypotheses}
For this work we will make six assertions :
\begin{enumerate}
  \item Initial probabilities are known.
  \item Only single atoms can move.
  \item Single atom (monomer) diffusion can be decomposed in two half steps :
    \begin{enumerate}
    \item A monomer is removed from the set of monomers.
    \item The taken atom is then placed randomly at the surface (i.e. on a nucleation site)
    \end{enumerate}
  \item The number $N$ of nucleation sites is very large : $1/N \gg 1
$  \item The mean free path of a diffusing atom is large in comparison of the distance of
    nucleation center.
  \item Finally, the size of a cluster is negligible in regard of the distance of
    nucleation center. This means that the capture probability for an atom does not
    depend of the size of the cluster already present on the center, and that condition
    5 will be fulfilled.
\end{enumerate}

\section{Result}
With these assumptions and some tedious calculations shown in \nameref{annexe1}, it comes the
following formula that give the final distribution after all monomers have diffused :

\begin{equation}
  \begin{aligned}
    \underline{P}_{n \ne 0} ~~  &=~~ e^{-x_a} \sum_{i=0}^{n} \frac{P_{n-i}{~} x_a^i }{i!}
      ~~ - ~~ e^{-x_a} \frac{x_a^n}{n!}\\
    ~\\
    \underline{P}_0 ~ &= ~ ( P_0 - 1 ) e^{-x_a} + 1\\
    ~\\
   & with\\
    ~\\
x_a ~ &= ~ \frac {P_1} {( 1 - P_0 )}
\end{aligned}
\label{result1} 
\end{equation}

where $P_n$ is the initial probability to have a cluster of size $n$ atoms and $\underline{P}_n$ is
the final probability to have a cluster of $n$ atoms. $x_a$ is the mean number of diffusions per
site to clean the surface of all single atoms, as properly explained in \nameref{annexe1}.

\section{Some remarks}

\subsection{remark 1:}
If only monomers are found at the beginning of the diffusion process, then $P_0 = 1-P_1$
implying that $x_a=1$. The mean number of diffusions per site to get rid of monomers is 1,
and this, whatever the value of $P_1$.

\subsection{remark 2:}

One expects that the statistic will not change much when the number of monomers
and the number of vacant sites are small. In these conditions, according to the formula
giving $x_a$: $x_a = \frac {P_1} {( 1 - P_0 )}$ , we expect only a few numbers of steps before
the disappearance of all the monomers. What happens for the probability $P_0$ is particularly
interesting since $P_0$ is easily measurable experimentally, using an STM for example,
(which is not necessarily true for other size classes). With $P_1$ and $P_0$ small,
$x_a$ is small, and one can make a Taylor expansion of the exponential term in the formula
giving the probability of having an empty site: $e^{-x_a} = 1 - x_a + \frac {1} {2} x_a^2 +...$.
Replacing $x_a$ with its expression depending on $P_0$ and $P_1$, one gets:

\[
e^{-x_a} \approx 1 + \frac {P_1} {P_0 - 1} + \frac {1}{2} \frac {P_1^2} {( P_0 - 1)^2}
\]

and finally :

\[
\underline{P}_0 \approx P_0 + P_1 - \frac {1}{2} P_1^2
\]

In the case of a deposit of more than 3 atoms per site, the Poisson distribution is such
that $\underline{P}_0$ will be very little different from $P_0 + P_1$ . It’s only for deposits
lower than 3 atoms per site that this statistic will clearly distinguish itself from the trivial
situation in which one simply removes the monomers without caring about anything else.

\subsection{remark 3 - Size histogram before coalescence:}

Let's go back to the case where only monomers are present: $P_0 = 1 - P_1$. As we have seen:
$x_a = 1$. Since only $P_0$ and $P_1$ are different from 0, the sum in formula
(\ref{result1}) is simplified, and we obtain:

\[
\begin{aligned}
  \underline{P}_{n \ne 0} &= \left [ \sum_{i=0}^{n-1}  \frac{P_{n-i}{~} x_a^i}{i!}
    + \frac{1}{n!}( P_0 - 1 ) x_a^n \right ] e^{-x_a}\\~\\
  &= \left [ P_1 \frac{x_a^{n-1}}{(n-1 )!}
    + \frac{1}{n!} ( P_0 - 1 ) x_a^n \right ] e^{-x_a}
\end{aligned}
\]

as  $x_a = 1$ and  $P_0 - 1 = -P_1$,

\[
\begin{aligned}
  \underline{P}_{n \ge 2} &= \left [ \sum_{i=0}^{n-1} \frac{P_{n-i} x_a^i}{i!}
      + \frac{1}{n!} ( P_0 - 1 ) x_a^n \right ] e^{-x_a} \\~\\
    &= \left [ \frac{P_1}{(n-1)!}  - \frac{P_1}{n!} \right ] \frac{1}{e} \\~\\
    &= \frac{1}{e} \frac{n-1}{n!}P_1
\end{aligned}
\]

\begin{equation}
\underline{P}_{n\ge2} = \frac{1}{e} \frac{(n-1 )}{n!} P_1
\label{result2a} 
\end{equation}

The probability of presence for sizes $n \ge 2$ is strictly proportional to $P_1$. Whatever
the value of $P_1$. Considering the sizes $n$ and $n'$ the ratio $\underline{P}_n /
\underline{P}_n'$ is a constant. The limit where $P_1$ tends towards 0 is interesting and
deserves to be interpreted physically: First, for $P_1 \ll 1$, the atoms arranged on the
surface are totally random and do not show any particular order: the underlying lattice
of nucleation sites appears continuous in the same way the granular structure of the
atoms fades away when considering a macroscopic object. Secondly, the successive diffusions
will randomly impact different monomers, and although the procedure described here starts from
a sequential process, it is in parallel that the individual atoms will diffuse and meet each
other. The meeting place is absolutely random and in fact, describes homogeneous nucleation
with a critical germ equal to 2. The different probabilities $\underline{P}_n$ give, ignoring
a multiplier factor, the quantity of cluster of size s just after nucleation and before a
possible coalescence.

$\underline{P}_n$ can be amplified by a factor $\alpha$ to ensure that
$\sum_{n=2}^\infty ( \alpha \underline{ P}_n ) = 1$. This factor is easy to calculate, indeed :

\[
\sum_{n=2}^\infty \underline{P}_n + \underline{P}_0
= 1 \Leftrightarrow \sum_{n=2}^\infty \underline{P}_n = 1 - \underline{P}_0 = \frac{1}{\alpha}
\]

as

\[
\underline{P}_0 = ( P_0 - 1 ) e^{-x_a} + 1 = 1 - \frac{1}{e} P_1
\]
one gets
\[
\alpha = \frac {e} {P_1}
\]
by including $\alpha$ in formula (\ref{result2a}), we finally get :

\begin{equation}
\Pi_{n\ge2} =  \frac{(n-1)}{n!}
\label{result2b} 
\end{equation}

$\Pi_n$ being the probabilities after normalization.
It is shown in  \nameref{annexe2} in the section \ref{preuve_convergence} that, as
expected, $\sum_{n=2}^\infty \Pi_n = 1$
Formula (\ref{result2b}) gives the size histogram in a system where only homogeneous nucleation
operates, with a critical germ equal to 2, and before a possible coalescence or eventual growth
in case monomers are injected (by evaporation-condensation for example) after the primordial
nucleation process. To reach such a statistic experimentally, it will be necessary, either to
make an evaporation of very short duration so that the nucleation has no time to really start
before the end of the evaporation process, deposit at low temperature followed by annealing to
allow the monomers to move, or then, in the gas phase, a very brutal cooling (i.e. in a
supersonic jet after laser ablation). It is also necessary, that the cluster sizes are negligible
compared to the characteristic distances in the system, to ensure a large average free
path and a capture probability independent of the size of growing clusters.\\
The average size of the clusters thus formed is easily calculated (see \ref{preuve_nombre_homogene}
in \nameref{annexe2}) :

\begin{equation}
  \overline{n} = \sum_{n=2}^\infty s\Pi_n = e
  \label{nombre_homogene}
\end{equation}

This makes it easy to test formula (\ref{result2b}): Indeed, if $N$ atoms have been deposited
on a given surface, one should expect to be able to observe only $N/e$ clusters, i.e. approximately
2.7 times less than the number of atoms present. It should be quite easy to verify this prediction
by choosing a suitable atom-substrate pair. The only constraint is that, since we are dealing
with homogeneous nucleation, the substrate must be free of defects.

\section{Confronting experience}
\label{confrontation}
The work described in reference \cite{ISI:000488835000007} shows the interest that can have the
knowledge of these formulas to characterize a studied system. In this paper, Düll et al. make a
platinum deposit on an h-BN/Rh(111) moiré. It is said that a Mono Layer (ML) of platinum corresponds
to 144 atoms per cell. They show a result obtained at 295K by STM imaging, after a deposit
of 0.005 ML at 295K. As written in the article: << $\sim$27\% of the h-BN pores are filled with Pt
clusters,... >>. They write also that, given the small size of the clusters, some of the Pt
clusters are expected to consist of just one single atom.\\

First, just after deposition, we would expect to obtain a Poisson distribution as all cells are
equivalent. 0.005 ML corresponds to an average occupancy of 0.72 atoms per site, and the expected
Poisson distribution is :

$P_0 = 0.4867$\\
$P_1 = 0.3504$\\
$P_2 = 0.1261$\\
$P_3 = 0.0302$\\
$P_4 = 0.0054$\\
$P_5 = 0.0008$

meaning that the number of occupied sites should be $1 - P_0 =  1 - 0.4867 = 0.51331$, that
means 51\%, far from the 27\% observed in the paper.

Considering monomer diffusion and time enough between deposition and STM observations,
we can apply the formulas given in the present work: these probabilities become
(after 0.6828 movements per site on average):

$P_0  :  0.4867  \to  0.7407$\\
$P_1  :  0.3504  \to  0.0$\\
$P_2  :  0.1261  \to  0.1241$\\
$P_3  :  0.0302  \to  0.0863$\\
$P_4  :  0.0054  \to  0.0351$\\
$P_5  :  0.0007  \to  0.0105$

The occupancy rate is now only $1 - 0.7407 = 0.2593$, or 26\%, which corresponds well to
the experimental observation. We can add that, when STM images have been recorded,
none of the cell was occupied by a single atom.

\section{Conclusion}
The size distribution exact law (\ref{result1}) after the monomer diffusion is deduced rigorously by
assuming some assumptions (germination centers on an array, mobility of the monomers). This formula
predicting the size statistic modification should be useful to all experimenters working with very
small deposits of atoms or molecules on surfaces wishing to best characterize their studied samples.
As an exact law, one can easily imagine that it could also be useful in other fields for which I have
no particular expertise. For example, I am thinking about the diffusion and coalescence of atoms
on powders, or about the nucleation in gel solutions. A consequence of this law is that it allows
predicting the size distribution after homogeneous nucleation by a simple and concise formula
(\ref{result2b}), which is a fundamental result that could be useful in many fields of physics:
laser ablation, very short duration deposits of atoms, nucleation of droplets in cloud
chambers, etc.

\section{Acknowledgments}
I want to thank my colleague Claude Henry, for fruitful discussions and valuable assistance to
find hided mistakes in the formulas.

\bigskip

\newpage


\newpage
\section{Annex 1 - Calculations}
\label{annexe1}

From \nameref{hypotheses} we can calculate exactly, diffusion by diffusion, the modification
of the size histogram.

We will note than the first half step of the diffusion affects only the amount of monomers and
empty sites, and the second step will affect all size classes.

\subsection{Enumeration of size’s classes and evolution of probabilities}
Let be an initial configuration where there is the probability $P_0$ to have an empty site,
the probability $P_1$ to have a site with a single atom, $P_2$ to have a site with a dimer… and
$P_n$ to have a site with a cluster of size n atoms. After a number $a$ of diffusions, the new
probabilities will be noted ${}^aP_0$, ${}^aP_1$, ${}^aP_2$, ${}^aP_3$,  ... Let’s see to
start, how $P_0$ transforms during a single diffusion.

\subsubsection{$P_0$}
Let $N_0$ be the number of empty sites before the diffusion.  $N_0 = NP_0$.
The first half step of the diffusion rises by 1 the number of empty sites :  $N_0 \to N_0’ = N_0 + 1$ .
The new probability $P_0’$ is then in a way that $N_0’ = NP_0’$ i.e.  $NP_0 + 1 = NP_0’$. It therefore
follows that $P_0’ = P_0 + 1/N$. Since now we will substitute $1/N$ by $\varepsilon$ to simplify
the notation.

\[
P_0' = P_0 + \varepsilon
\]

The second half step will decrease the amount of empty sites, and this, according to the probability
to find an empty site : $N_0' \to N_0'' = N_0' - (1 \times P_0')$. In the same way that done just above,
we can verify that, if $N_0'' = NP_0''$ then
$P_0'' = P_0' - \varepsilon P_0' = ( 1 - \varepsilon )P_0' = (1 - \varepsilon )(P_0+\varepsilon )$.

\[
P_0'' = P_0' - P_0' \varepsilon
\]

At the end of the whole diffusion :

\[
{}^1P_0 = (P_0 +  \varepsilon)(1 - \varepsilon)
\]

However, it is interesting at this point to raise a little bit the difficulty of the game by assuming
that each diffusion can rise from a monomer formerly present on the surface or from an atom coming
from outside (gas phase during the deposition for example). Then the first half step will not subtract
$\varepsilon$ from $P_1$ but $\beta$ ($0 \le \beta \le \varepsilon$). This complication will
nevertheless lead to the solution expected for our initial problem when setting $\beta = \varepsilon$,
and allow us to check if we are wrong or not: indeed, by setting $\beta = 0$, one should expect to
rediscover the Poisson distribution. We can now rewrite the probability after the diffusion :

\[
{}^1P_0 = (P_0 + \beta)(1 - \varepsilon)
\]

Trivially, the upper formula can be generalized to the recurrence relation :

\begin{equation}
{}^{a+1}{P_0} = ({}^{a}P_0 + \beta)(1-\varepsilon)
\label{evolution_P0_sale} 
\end{equation}

\subsubsection{$P_1$}
It’s a little bit more complicated to address $P_1$. Indeed, the successive values of $P_1$ will not only
depend on the value of $P_1$ at the former stage, but also on the former value of $P_0$. After the first
half step, one finds:

\[
P_0' = P_0 + \beta ~~ and ~~ P_1' = P_1 - \beta
\]

If the moving atom is placed on an empty site (with probability $P_0'$) $P_1'' =  P_1' + \varepsilon$,
and if the moving atom is placed on a site with a single atom (with probability $P_0'$),
$P_1'' =  P_1' - \varepsilon$. It follows that :

\[
{}^1P_1 = (P_1 - \beta)(1- \varepsilon) + \varepsilon (P_0 + \beta)
\]

And again, we can very easily deduce the recurrence relation:

\begin{equation}
{}^{a+1}P_1 = ({}^{a}P_1 - \beta)(1- \varepsilon) + \varepsilon ({}^{a}P_0 + \beta)
\label{evolution_P1_sale} 
\end{equation}

\subsubsection{$P_2$}
The value of $P_2$ is not modified by the first half step but changed if during the second stage
of the diffusion the atom is deposited on a preexisting monomer or a preexisting dimer. If the
atom is dropped on a site with one monomer $P_2'' = P_2 + \varepsilon$  , and if it is dropped
on a site with a dimer $P_2'' = P_2 - \varepsilon$  . Taking into account the probabilities of
having a monomer or a dimer it follows the recurrence relation:

\begin{equation}
{}^{a+1}P_2  = {}^{a}P_2 ( 1- \varepsilon) + \varepsilon({}^{a}P_1  - \beta)
\label{evolution_P2_sale} 
\end{equation}

\subsubsection{$P_3$}
In the same way, there are two means to change the probability of having a trimer: Depositing the
diffusing atom on a site with a trimer ($-\varepsilon$), or depositing the atom on a dimer
($+\varepsilon$). The associated recurrent formula raises:

\begin{equation}
{}^{a+1}P_3  =  {}^{a}P_3( 1- \varepsilon)  + \varepsilon {}^{a}P_2
\label{evolution_P3} 
\end{equation}

\subsubsection{$P_n$}
From s=3, we can notice that the result doesn't any-more depend on the first step of the diffusion
but only  on the former values of $P_n$ and $P_{n-1}$. So for $s \ge 3$:

\begin{equation}
{}^{a+1}P_n  ~ = ~ {}^{a}P_n ( 1- \varepsilon) + ~ \varepsilon ~ {}^{a}P_{n-1} 
\label{evolution_P3} 
\end{equation}

\subsection{Iterations}
\label{iterations}

To obtain the final statistics of site occupation, one has to iterate the elementary process
and get an analytic expression of ${}^{a}P_n$ . The expression of  ${}^{a}P_1$ is of main importance
because the diffusion process must stop when  ${}^{a}P_1=0$. This will be examined in the section
\nameref{interruption}.\\
Now we are going to concentrate on obtaining the expressions of ${}^{a}P_n$  depending on
$a$, $\varepsilon$, $\beta$, and of the initial values of $P_n$ . Notice that as the number $N$ of sites
is very large,  $\varepsilon$ and $\beta$ are tiny. The number $a$ of iterations is proportional
to $N$, as increasing by a factor $A$ the number of sites will automatically increase by the same
factor $A$ the number of elementary processes to reach the same situation. The product
$a \times \varepsilon$ is in fact the mean number of diffusion per site and I will note $x$ this
number when the discrete formulas will be extrapolated to continuous formula.\\
Second-order terms implying $\varepsilon^2$ , $\varepsilon\beta$ or $\beta^2$ are negligible
compared to 1, except if associated with the number $a$ of steps, and, of course, 1
is negligible compared to $a$.

\subsubsection{$P_0$}
After neglecting what has to be neglected in formula (\ref{evolution_P0_sale}), we have :

\begin{equation}
{}^{a+1}P_0  ~ = ~ {}^{a}P_0 ( 1- \varepsilon) + ~ \beta
\label{evolution_P0} 
\end{equation}

Let us try to find a kind of regularity to the successive expressions of ${}^{a}P_0$ :

\[
\begin{aligned}
P_0 {}^{1} &= P_0 (1 - \varepsilon ) + \beta \\ 
P_0 {}^{2} &= P_0 (1 - \varepsilon )^2 + \beta (1 - \varepsilon ) + \beta \\ 
P_0 {}^{3} &= P_0 (1 - \varepsilon )^3 + \beta (1 - \varepsilon )^2 + \beta (1 - \varepsilon ) + \beta
\end{aligned}
\]

A form seems to emerge:

\[
{}^{a}P_0  ~ = ~ {}^{0}P_0  (1 - \varepsilon )^a + \beta \sum_{i=0}^{a-1} (1 - \varepsilon )^i
\]

It is easy to calculate the sum:

\[
\sum_{i=0}^{a-1} (1 - \varepsilon )^i ~ = ~ \frac{(1 - \varepsilon )^a - 1 }{(1 - \varepsilon ) - 1 }
~ = ~ - \frac{1}{\varepsilon} \left [ (1 - \varepsilon )^a - 1 \right ] 
\]

And then, as ${}^0P_0$ is nothing else than $P_0$:

\[
{}^{a}P_0  ~ = ~ P_0 (1 - \varepsilon )^a + \frac{\beta}{\varepsilon} [ 1 - ( 1 - \varepsilon )^a ]
\]

\begin{equation}
{}^{a}P_0  ~ = ~ (P_0 - \frac{\beta}{\varepsilon} )( 1 - \varepsilon )^a ~ + ~ \frac{\beta}{\varepsilon}
\label{P0d} 
\end{equation}
The proof of this formula can be found in \nameref{annexe2} -  \nameref{preuve_P0}.
This discrete formula (\ref{P0d}) can be expressed in a continuous formula (knowing that
$(1-\varepsilon)^a = e^{-a\varepsilon}$, when $-\varepsilon \ll 1$ ) :

\begin{equation}
P_0(x) ~ = ~ (P_0 - \frac{\beta}{\varepsilon}  )e^{-x} ~ + ~ \frac{\beta}{\varepsilon}
\label{P0c} 
\end{equation}

\subsubsection{$P_1$}
As for $P_0$, we will neglect the quadratic terms of $\varepsilon$ in the recurrent formula
(\ref{evolution_P1_sale}) for $P_1$, Leading to:

\begin{equation}
{}^{a+1}P_1  ~ = ~  {}^{a}P_1( 1- \varepsilon) ~ + ~ \varepsilon ~ {}^{a}P_0  ~ - ~ \beta
\label{evolution_P1} 
\end{equation}

After iterations and search for regularities, it turns out that:

\[
  {}^{a}{P_1} ~ = ~ P_1( 1- \varepsilon)^a ~ + ~ \varepsilon \sum_{i=0}^{a-1}
  [ {}^{a-1-i}{P_0} ( 1- \varepsilon)^i] ~ - ~ \beta \sum_{i=0}^{a-1} ( 1- \varepsilon)^i
\]

Witch can be simplified into:

\begin{equation}
  {}^{a}{P_1} ~ = ~ P_1 ( 1- \varepsilon)^a ~ + ~(P_0 - \frac {\beta}
  {\varepsilon}) a \varepsilon ( 1- \varepsilon )^{a-1}
\label{P1d} 
\end{equation}

The proof of this formula can be found in \nameref{annexe2} subsection  \nameref{preuve_P1}\\
As for $P_0$, we can express a continuous form of this formula :

\begin{equation}
P_1(x) ~ = ~ ( P_1 ~ + ~ x (P_0 - \frac{\beta}{\varepsilon}) ) e^{-x}
\label{P1c} 
\end{equation}

\subsubsection{$P_2$}
Once the quadratic terms of formula (\ref{evolution_P2_sale}) removed we have :

\begin{equation}
{}^{a+1}P_2  ~ = ~  {}^{a}P_2( 1- \varepsilon) ~ + ~ \varepsilon ~ {}^{a}P_1
\label{evolution_P2} 
\end{equation}
and, by the mean already used, the solutions is :
\[
  {}^{a}{P_2} ~ = ~ P_2( 1- \varepsilon)^a ~
  + ~ \varepsilon \sum_{i=0}^{a-1} [ {}^{a-1-i} {P_1} ( 1- \varepsilon)^i]
\]
Leading to :
\begin{equation}
{}^{a}{P_2} = P_2 (1- \varepsilon)^a + P_1 a \varepsilon ( 1- \varepsilon )^{a-1}
  + (P_0 - \frac{\beta}{\varepsilon}) \frac {a(a-1)}{2} \varepsilon^2 ( 1- \varepsilon )^{a-2}
\label{P2d} 
\end{equation}
The proof of this formula can be found in \nameref{annexe2} subsection  \nameref{preuve_P2}.
and the associated continuous formula takes the form:

\begin{equation}
P_2 (x) = \left [ P_2 + P_1 x + (P_0 - \frac {\beta}{\varepsilon}) \frac{x^2}{2} \right ] e^{-x}
\label{P2c} 
\end{equation}

\subsubsection{$P_3$}
The recurrence relation for $P_3$ is the same as for $P_2$, and one could expect that all can
be deduced easily from here, however, the expressions becomes more and more complicated :
\begin{equation}
{}^{a+1}{P_3} = {}^{a}{P_3} (1- \varepsilon) + \varepsilon {}^{a}{P_2}
\label{evolution_P3b} 
\end{equation}

and so :
\begin{equation}
{}^{a}{P_3} = P_3( 1- \varepsilon)^a + \varepsilon \sum_{i=0}^{a-1} [ {}^{a-1-i}{P_2} ( 1- \varepsilon)^i]
\label{evolution_P3c} 
\end{equation}
To solve this, we need to sum the square of integers from $0$ to $(a-2)$ \cite{Dostor}. And finally, we get :

\begin{equation}
  {}^{a} {P_3} = P_3 ( 1- \varepsilon)^a + P_2 a\varepsilon ( 1- \varepsilon )^{a-1}
  + \frac {1} {2} P_1 {a (a-1)} \varepsilon^2 ( 1- \varepsilon )^{a-2}
  + \frac {1} {6} (P_0 - \frac {\beta} {\varepsilon} ){(a-1)[(a-1)^2 - 1 ]} \varepsilon^3
  ( 1- \varepsilon )^{a-3}
\label{P3d} 
\end{equation}
The proof of this formula can be found in \nameref{annexe2} subsection   \nameref{preuve_P3}.
The discrete formula (17) leads to the following continuous formula:

\begin{equation}
  P_3(x) = \left [ P_3 + P_2 x + \frac {1} {2} P_1 x^2
    + \frac {1} {6} (P_0 - \frac {\beta} {\varepsilon})x^3 \right ] e^{-x}
\label{P3c} 
\end{equation}

\subsection{Short discussion}
Trying to generalize what is obtained for $P_3$, it seems that the beginning of $P_n$ is :

\[
\begin{aligned}
  {}^{a}{P_n} = P_n ( 1- \varepsilon)^a + P_{n-1} a\varepsilon ( 1- \varepsilon )^{a-1}
  + \frac {1} {2} P_{n-2} {a (a-1)} \varepsilon^2 ( 1- \varepsilon )^{a-2} \\
  + \frac {1} {6} P_{n-3}(a-1)[(a-1)^2 - 1 ] \varepsilon^3 ( 1- \varepsilon )^{a-3}
  + ... + K (P_{0} - \frac {\beta} {\varepsilon} )
\end{aligned}
\]

\[
\begin{aligned}
  {}^{a}{P_n} = P_n( 1- \varepsilon)^a + K_1 P_{n-1}\varepsilon( 1- \varepsilon )^{a-1}
  + K_2 P_{n-2} \varepsilon^2 ( 1- \varepsilon )^{a-2} \\ + K_3 P_{n-3} \varepsilon^3 ( 1- \varepsilon )^{a-3}
  + ... + K_{n} (P_{0} - \frac {\beta} {\varepsilon} )
\end{aligned}
\]

Every coefficient $K_m$ associated with $P_{s-m}$ are coming from the sums found in the generic recurrence
formula for $P_{n-m-1}$:

\[
{}^{a}{P_n} = P_n( 1- \varepsilon)^a + \varepsilon \sum_{i=0}^{a-1} [ {}^{a-1-i}P_{n-1} ( 1- \varepsilon)^i]
\]

This means that to obtain $K_m$, we have to calculate the sum of the sum of the sum...
m times of something depending on $P_0$ , leading to calculate the sum of powers of $a$ at every
steep higher.
Ignoring all that can be neglected, the successive terms for $P_{n-m}$ look like $\alpha(a\varepsilon)^{m}$.
The problem being to determine $\alpha$ : Indeed, it does not exist a simple formula for the sum
of powers $\sum_{i=0}^N i^k$. We can use the Von Staudt formula \cite{VonStaudt} $\sum_{i=0}^{N} i^k = N^k
+ \sum_{i=0}^{k} \left [ \frac {B_i k!} { i! (k-i+1)! } N^{ k-i+1 } \right ]$,  $B_i$ being the
Bernoulli numbers : $B_0 = 1$ ; $B_1 = -1/2$ ; $B_2 = 1/6$ ; $B_3 = 0$ ; $B_4 = -1/30$ ... or
the Faulhaber formula \cite{ISI:A1993LQ84200023}, which also needs the Bernoulli numbers and is not simpler.\\
However, the case where $\beta = 0$ suggests that $\alpha$ should be equal to $m!$ , and
we would have the following continuous version for the sizes different from 0 :

\begin{equation}
  P_{n \ne 0} (x) = \left [ \sum_{i=1}^{n-1}  \frac {P_{n-i} x^i}{ i!}
    + \frac{1}{n!}( P_0 - \frac {\beta} {\varepsilon} )x^n \right ] e^{-x}
\label{Psc1} 
\end{equation}

\subsection{"Hybrid" proof of formula (\ref{Psc1})}
By "hybrid" I mean that instead of making iterations of the recurrence formula injecting
the discrete and complicated probability expression for the previous size, I will use the
continuous form obtained. Let us return to the procedure that allowed us to deduce $P_3$
, and use the continuous formula (\ref{P2c}) instead of the discrete one ((\ref{P2d}).\\

On one hand, we have :

\[
  {}^{a+1}{P_3} = {}^{a}{P_3} (1- \varepsilon) + \varepsilon {}^{a}{P_2}
\]

that leads to:

\[
{}^{a}{P_3} = P_3( 1- \varepsilon)^a + \varepsilon \sum_{i=0}^{a-1} [ {}^{a-1-i}{P_2} ( 1- \varepsilon)^i]
\]

and on the other hand:

\[
P_2 (x) = \left [ P_2 + P_1 x + (P_0 - \frac{\beta}{\varepsilon}) \frac{x^2}{2} \right ] e^{-x}
\]

Replacing $x$ by $\varepsilon a$, this formula transforms to :

\[
  {}^{a}{P_2} = \left [ P_2 + P_1 \varepsilon a
    + \frac{1}{2} (P_0 - \frac{\beta}{\varepsilon})\varepsilon^2 a^2 \right ] (1 - \varepsilon)^a
\]

that is, in fact, the discrete formula (\ref{P2d}) free from all negligible terms.\\
Let’s focus on the sum : $\sum_{i=0}^{a-1} [ {}^{a-1-i}{P_2} ( 1- \varepsilon)^i \varepsilon ]$.
Inverting the order of the terms will simplify the understanding of the meaning of this sum.

\[
\sum_{i=0}^{a-1} [ {}^{a-1-i} {P_2} ( 1- \varepsilon)^i \varepsilon ]
= \sum_{i=a-1}^{0} [ {}^{a-1-i} {P_2} ( 1- \varepsilon)^i \varepsilon ]
= \sum_{j=0}^{a-1} [ {}^{j} {P_2} ( 1- \varepsilon)^{a-1-j} \varepsilon ]
\]

I leave $\varepsilon$ inside the sum because it will have its importance.

\[
\begin{aligned}
  {}^{j}{P_2} ( 1- \varepsilon)^i &= \left [ P_2 + P_1 \varepsilon j
    + \frac {1} {2} (P_0 - \frac {\beta} {\varepsilon})\varepsilon^2 j^2 \right ]
  (1 - \varepsilon)^j \times (1 - \varepsilon)^{(a-1-j)} \\~\\
  &= \left [ P_2 + P_1 \varepsilon j
    + \frac {1} {2} (P_0 - \frac {\beta} {\varepsilon})\varepsilon^2 j^2 \right ] (1 - \varepsilon)^{a-1}
\end{aligned}
\]

when $\varepsilon$  tends towards 0, $(1-\varepsilon)^{a-1} =(1-\varepsilon)^{a} $ and finally :

\[
  {}^j{P_2} ( 1- \varepsilon)^{a-1-j} = \left [ P_2 + P_1 \varepsilon j
    + \frac{1}{2} (P_0 - \frac {\beta} {\varepsilon})\varepsilon^2 j^2 \right ] e^{-\varepsilon a}
\]

We will notice, once again when  $\varepsilon$  tends towards 0, that:

\[
\sum_{j=0}^{a-1} \left [ \left [ P_2 + P_1 \varepsilon j
    + \frac {1} {2} (P_0 - \frac {\beta} {\varepsilon})\varepsilon^2 j^2 \right ]
  e^{-\varepsilon a} \times \varepsilon \right ]
= \int_{0}^{x} \left [ P_2 + P_1 \alpha + \frac{1}{2} (P_0 - \frac {\beta} {\varepsilon})
  \alpha^2 \right ] e^{-x} d\alpha
\]

And the continuous form for $Py_3$ is directly deduced, avoiding the tedious calculus of
the discrete formulas:

\[
\begin{aligned}
  P_3 (x) &= P_3 e^{-x} + \int_0^x \left [ P_2 + P_1 \alpha
    + \frac{1}{2} (P_0 - \frac{\beta}{\varepsilon})\alpha^2 \right ] e^{-x} d\alpha \\
  &= P_3 e^{-x} + e^{-x} \int_0^x P_2( \alpha )d\alpha \\
  &= \left [ P_3 + P_2 x + \frac {1} {2} P_1 x^2 + \frac {1} {6}
    (P_0 - \frac{\beta}{\varepsilon}) x^3 \right ] e^{-x}
\end{aligned}
\]

and we can even write that :

\[
P_3 (x) = P_3 e^{-x} + e^{-x} \int_0^x \left [P_2 + \int_0^\gamma  \left [ P_1
    + \int_0^\delta \left [ (P_0 - \frac {\beta} {\varepsilon}) \right ]
    d\alpha \right ] d\delta \right ] d\gamma
\]

Considering the probability $P_n$ for the size s requires an integration that raises the
power of $x$ and brings out, as expected, the factorial of the number of successive integrations.

\subsection{Interrupting the diffusion}
\label{interruption}
As explained at the beginning of \nameref{iterations}, the diffusion will stop after $x_a$ movements
per site, when no more monomers are present. One has to solve the following equation :
\[
P_1 ( x ) = ( P_1 + x ( P_0 - \frac {\beta} {\varepsilon} ) ) e^{-x} = 0
\]
If $xa$ exists, it is such that $P_1 + x_i ( P_0 - \frac {\beta} {\varepsilon} ) = 0$
is to say :
\[
x_a = \frac {P_1} {\frac {\beta} {\varepsilon} - P_0 }
\]

Let us remember that here, we are interested by the particular case where $\beta=\varepsilon$.
Finally the new distribution, after the monomer diffusion process compleeted, is expressed in
the following concise form [equation (\ref{result1})]:

\begin{equation}
  \begin{aligned}
    \underline{P}_{n \ne 0} = P_{n \ne 0} ( x_a )
    &= \left [ \sum_{i=0}^{n-1} \frac{P_{n-i}{~} x_a^i }{i!}
      + \frac{1}{n!} ( P_0 - 1 ) x_a^n \right ] e^{-x_a}\\
    ~\\
    \underline{P}_0 = P_0 ( x_a ) &= ( P_0 - 1 ) e^{-x_a} + 1\\
    ~\\
   & with\\
    ~\\
x_a &= \frac {P_1} {( 1 - P_0 )}
\end{aligned}
\end{equation}

\subsection{Some verifications}

\subsubsection{Sum of probabilities}
We have to check, at least in specific cases, that the sum of the probabilities (\ref{result1})
is equal to 1. Let's check what happens if all stating probabilities are equal to zero
except $P_1 = 1$ :
To lighten the notation, let's replace $\beta/\varepsilon$ with $\gamma$.
\[
\begin{aligned}
P_0 ( x ) e^x &= \gamma e^x - \gamma\\
P_1 ( x ) e^{x} &= 1 - \gamma x\\
P_2 ( x ) e^{x} &= x - \frac {1} {2} \gamma x^2\\
P_3 ( x ) e^{x} &= \frac {1} {2} x^2 - \frac {1} {6} \gamma x^3\\
P_4 ( x ) e^{x} &= \frac {1} {6} x^3 - \frac {1} {24} \gamma x^4\\
P_{n-1} ( x ) e^{x} &= \frac {1}{n-2)!} x^{n-2} - \frac{1}{n-1)!} \gamma x^{n-1}\\
P_n ( x ) e^{x} &= \frac{1}{n-1)!} x^{n-1} - \frac {1} {n!} \gamma x^n\\
\end{aligned}
\]
\[
\begin{aligned}
\sum_{n=0}^\infty P_n( x )e^{x} &= \gamma e^x + ( 1-\gamma ) + x ( 1-\gamma )
+ \frac{1}{2} x^2 ( 1-\gamma ) + \frac{1}{3!} x^3 ( 1-\gamma )
+ ... + \frac{1}{n!} x^n ( 1-\gamma ) + ... \\
&= \gamma e^x + ( 1-\gamma ) \sum_{n=0}^\infty \left ( \frac {x^n}{n!} \right ) \\
&= \gamma e^x + ( 1-\gamma ) e^x = e^x
\end{aligned}
\]
\[
\sum_{n=0}^\infty P_n ( x ) e^{x} = e^x \Leftrightarrow \sum_{n=0}^\infty P_n ( x ) =1
\]

\subsubsection{Poisson distribution}
By taking $\beta=0$, $P_0=1$, and of course all $P_{n\ne0}=0$, we expect to find the
values given by the Poisson Distribution. 
\[
\begin{aligned}
P_{n \ne 0} ( x ) &= \left [ \sum_{i=0}^{n-1}\frac{P_{n-i}{~} x^i}{i!}
  + \frac{1}{n!} ( P_0 - \frac{\beta}{\varepsilon} ) x^n \right ] e^{-x} \\
&= \left [ \sum_{i=0}^{n-1}\frac{0 \times x^i}{n!} 
  + \frac{1}{n!} 1 \times x^n \right ] e^{-x} \\
&= \frac {x^n}{n!} e^{-x}
\end{aligned}
\]
\[
P_0 ( x ) = ( P_0 - \frac{\beta}{\varepsilon} ) e^{-x} + \frac{\beta}{\varepsilon} = e^{-x}
\]
which matches exactly the Poisson distribution:
\[
P ( x,n ) = \frac{ x^n}{ n! } e^{-x}
\]

\subsubsection{Evolution of Poisson distribution}
Again with $\beta=0$, we can have a Poisson distribution with the corresponding probabilities
$P_0$, $P_1$, $P_2$,... as the starting point, and check how transforms these probabilities
during an extra deposit of atoms. As an example, I will focus on the evolution of $P_3$:

\[
P_3 ( x ) = \left [ P_3 + P_2 x + \frac{1}{2} P_1 x^2 + \frac{1}{6} P_0 x^3 \right ] e^{-x}
\]

The Poisson distribution for a mean number y of atoms per site is:

\[
\begin{aligned}
  P_0 &= e^{-y} \\
  P_1 &= y e^{-y} \\
  P_2 &= \frac {1} {2} y^2 e^{-y} \\
  P_3 &= \frac {1} {6} y^3 e^{-y} \\
\end{aligned}
\]

After an extra deposition of a mean number of x atoms per site is $P_3$ becomes :

\[
\begin{aligned}
  P_3 ( x ) &= \left [ \frac {1} {6} y^3 e^{-y} + \frac {1} {2} y^2 e^{-y} x
    + \frac {1} {2} y e^{-y} x^2 + \frac {1} {6} e^{-y} x^3 \right ] e^{-x} \\
  &= \left [ \frac {1} {6} y^3 + \frac {1} {2} y^2 x + \frac {1} {2} y x^2
    + \frac {1} {6} x^3 \right ] e^{-x} e^{-y} \\
  &= \frac {1} {6} \left [ y^3 + 3 y^2 x + 3 y x^2 + x^3 \right ] e^{- ( x+y ) } \\
  &= \frac {1} {6} ( x+y ) ^3 e^{- ( x+y ) }
\end{aligned}
\]

that matches the expected probability for trimers with a mean number of $x+y$ atoms par site.
\newpage
\section{Annex 2 - Supplementary material}
\label{annexe2}

Proof of the discrete formulas. All the formulas are proved by induction.

\subsection{Proof of the formula for $P_0$:}
\label{preuve_P0}
The expression (\ref{P0d}) is easily proven. We want to check that:
\[
  {}^{a}{P_0} = P_0 ( 1 - \varepsilon )^a - \frac {\beta} {\varepsilon} ( 1 - \varepsilon )^a
  + \frac {\beta} {\varepsilon}
\]
and we know the following recurrence expression:
\[
{}^{a+1} {P_0} = {}^{a}{P_0} ( 1 - \varepsilon ) + \beta
\]
Few lines allows us to calculate ${}^{a+1}P_0$ and check that the result is consistent
with the formula (\ref{P0d}) :

\[
\begin{aligned}
  {}^{a} {P_0} ( 1 - \varepsilon ) + \beta
  &= \left [ P_0 ( 1 - \varepsilon ) ^a - \frac {\beta} {\varepsilon} ( 1 - \varepsilon ) ^a
    + \frac {\beta} {\varepsilon} \right ] ( 1 - \varepsilon ) + \beta \\
  &= P_0 ( 1 - \varepsilon ) ^{a+1} - \frac {\beta} {\varepsilon} ( 1 - \varepsilon ) ^{a+1}
  + \frac {\beta} {\varepsilon} ( 1 - \varepsilon ) + \beta \\
  &= P_0 ( 1 - \varepsilon ) ^{a+1} - \frac {\beta} {\varepsilon} ( 1 - \varepsilon ) ^{a+1}
  + \frac {\beta} {\varepsilon} - \beta + \beta \\
  &= P_0 ( 1 - \varepsilon ) ^{a+1} - \frac {\beta} {\varepsilon} ( 1 - \varepsilon ) ^{a+1}
  + \frac {\beta} {\varepsilon} = {}^{a+1} {P_0}
\end{aligned}
\]
and obviously :
\[
{}^{0} {P_0} = ( P_0 - 1 ) ( 1 - \varepsilon ) ^0 + 1 = P_0 - 1 + 1 = P_0
\]
The accuracy of the formula (\ref{P0d}) is then proved

\subsection{Proof of the formula for $P_1$:}
\label{preuve_P1}
We have to check that the formula (\ref{P1d}) is correct. I remind this formula here :
\[
  {}^{a}{P_1} = P_1 ( 1- \varepsilon )^a
  + ( P_0 - \frac {\beta} {\varepsilon} ) a \varepsilon ( 1- \varepsilon )^{a-1}
\]
and we know that (formula (\ref{evolution_P1}) ):
\[
{}^{a+1}{P_1} = {}^{a}{P_1} ( 1- \varepsilon ) + \varepsilon {}^{a}{P_0} - \beta
\]
The formula follows correctly the recurrence law as we can check in the following calculus :

\[
\begin{aligned}
  {}^{a} {P_1} ( 1 - \varepsilon ) + \varepsilon {}^{a} {P_0} - \beta 
  &= P_1 ( 1 - \varepsilon )^{a+1} + ( P_0 - \frac {\beta} {\varepsilon} ) a \varepsilon
  ( 1- \varepsilon )^{a} + \varepsilon {}^{a} {P_0} - \beta \\
  &= P_1 ( 1 - \varepsilon )^{a+1} + ( P_0 - \frac {\beta} {\varepsilon} ) a \varepsilon
  ( 1- \varepsilon )^{a} + \varepsilon \left [ ( P_0 - \frac {\beta} {\varepsilon} )
    ( 1 - \varepsilon )^a + \frac {\beta} {\varepsilon} \right ] - \beta \\
  &= P_1 ( 1 - \varepsilon )^{a+1} + ( P_0 - \frac {\beta} {\varepsilon} ) a \varepsilon
  ( 1- \varepsilon )^{a} + \varepsilon ( P_0 - \frac {\beta} {\varepsilon} )
  ( 1 - \varepsilon )^a \\
  &= P_1 ( 1 - \varepsilon )^{a+1} + ( P_0 - \frac {\beta} {\varepsilon} )
  ( a+1 ) \varepsilon ( 1- \varepsilon )^{a} = {}^{a+1}{P_1}
\end{aligned}
\]
and we can check easily that :

\[
  {}^{0} {P_1} = P_1 ( 1- \varepsilon )^0 + ( P_0 - \frac {\beta} {\varepsilon} ) \times 0 \times
  \varepsilon ( 1- \varepsilon )^{-1} = P_1
\]
that definitively proofs the accuracy of the formula for(\ref{P1d}) $P_1$.

\subsection{Proof of the formula for $P_2$:}
\label{preuve_P2}
In one hand we have ${}^{a+1} {P_2} = P_2 {}^{a} { } ( 1- \varepsilon ) + \varepsilon {}^{a} {P_1}$
and ${}^{a} {P_1} = P_1 ( 1- \varepsilon )^a + ( P_0 - \frac {\beta} {\varepsilon} ) a
\varepsilon ( 1- \varepsilon )^{a-1}$ and we have to demonstrate that the following formula is correct :
\[
  {}^{a} {P_2} = P_2 ( 1- \varepsilon )^a + P_1 a\varepsilon ( 1- \varepsilon )^{a-1}
  + ( P_0 - \frac {\beta} {\varepsilon} ) \frac{a ( a-1 )}{2} \varepsilon^2 ( 1- \varepsilon )^{a-2}
\]
Let’s calculate a little bit :

\[
\begin{aligned}
  {}^{a}{P_2} ( 1- \varepsilon ) + \varepsilon {}^{a} {P_1}
  &= \left [ P_2 ( 1- \varepsilon )^a + P_1 a\varepsilon ( 1- \varepsilon )^{a-1}
    + ( P_0 - \frac {\beta} {\varepsilon} ) \frac{a ( a-1 )}{2} \varepsilon^2
    ( 1- \varepsilon )^{a-2} \right ] ( 1-\varepsilon )
  + \varepsilon{~} {}^{a}{P_1} \\~ \\~ \\
  &= P_2 ( 1- \varepsilon )^{a+1} + P_1 a\varepsilon ( 1- \varepsilon )^{a}
  + ( P_0 - \frac {\beta} {\varepsilon} ) \frac{a ( a-1 )} {2} \varepsilon^2
  ( 1- \varepsilon )^{a-1} \\
  &+ \varepsilon \left [ P_1 ( 1- \varepsilon )^a
    + ( P_0 - \frac {\beta} {\varepsilon} ) a \varepsilon ( 1- \varepsilon )^{a-1} \right ] \\~ \\~ \\
  &= P_2 ( 1- \varepsilon )^{a+1} + P_1 a\varepsilon ( 1- \varepsilon )^{a}
  + ( P_0 - \frac {\beta} {\varepsilon} )  \frac{a ( a-1 )}{2} \varepsilon^2 ( 1- \varepsilon )^{a-1} \\
  &+ P_1 \varepsilon ( 1- \varepsilon )^a + ( P_0 - \frac {\beta} {\varepsilon} ) a \varepsilon^2
  ( 1- \varepsilon )^{a-1} \\ ~\\~ \\
  &= P_2 ( 1- \varepsilon )^{a+1} + P_1 ( a+1 ) \varepsilon ( 1- \varepsilon )^{a}
  + ( P_0 - \frac {\beta} {\varepsilon} ) \varepsilon^2 ( 1- \varepsilon )^{a-1}
  \left [ \frac{a ( a-1 )} {2} + a \right ]
\end{aligned}
\]
since
\[
\frac{a(a-1)}{2} + a = \frac{( a+1 )a} {2}
\]
hence:
\[
{}^{a}{P_2} ( 1- \varepsilon ) + \varepsilon {}^{a}{P_1} = {}^{a+1}{P_2}
\]
And finally the particular case $a=0$ leads to ${}^0P_2 = P_2$ that demonstrates the correctness
of the formula.

\subsection{Proof of the formula for $P_3$:}
\label{preuve_P3}
We start with 
\[
{}^{a+1}{P_3} = {}^{a}{P_3} ( 1- \varepsilon ) + \varepsilon {}^{a}{ P_2}
\]
and
\[
  {}^{a} {P_2} = P_2 ( 1- \varepsilon )^a + P_1 a\varepsilon ( 1- \varepsilon )^{a-1}
  + ( P_0 - \frac {\beta} {\varepsilon} ) \frac{a(a-1)}{2} \varepsilon^2 ( 1- \varepsilon )^{a-2}
\]
, and we have to check that the following formula is correct :
\[
  {}^{a} {P_3} = P_3 ( 1- \varepsilon )^a + P_2 a\varepsilon ( 1- \varepsilon )^{a-1}
  + \frac{1}{2} P_1 {a ( a-1 ) } \varepsilon^2 ( 1- \varepsilon )^{a-2}
  + \frac{1}{6} ( P_0 - \frac {\beta} {\varepsilon} ) { ( a-1 ) [ ( a-1 )^2 - 1 ]}
  \varepsilon^3 ( 1- \varepsilon )^{a-3}
\]

\fontsize{9}{12}\selectfont
\[
\begin{aligned}
  &{}^{a}{P_3} ( 1- \varepsilon ) + \varepsilon {}^{a}{ P_2}\\
  &= \left [ P_3 ( 1- \varepsilon )^a + P_2 a\varepsilon ( 1- \varepsilon )^{a-1}
    + \frac{1}{2} P_1 {a ( a-1 ) } \varepsilon^2 ( 1- \varepsilon )^{a-2} 
     + \frac{1}{6} ( P_0 - \frac{\beta}{\varepsilon} ) { ( a-1 ) [ ( a-1 )^2 - 1 ]}
    \varepsilon^3 ( 1- \varepsilon )^{a-3} \right ] ( 1-\varepsilon )
  + \varepsilon {~}^{a}{ P_2} \\~\\~\\
  &= P_3 ( 1- \varepsilon )^{a+1} + P_2 a\varepsilon ( 1- \varepsilon )^{a}
  + \frac{1}{2} P_1 {a ( a-1 ) } \varepsilon^2 ( 1- \varepsilon )^{a-1}
  + \frac{1}{6} ( P_0 - \frac{\beta}{\varepsilon} ) { ( a-1 ) [ ( a-1 )^2 - 1 ]}
  \varepsilon^3 ( 1- \varepsilon )^{a-2} \\
  &+ \varepsilon \left [ P_2 ( 1- \varepsilon )^a + P_1 a\varepsilon ( 1- \varepsilon )^{a-1}
    + ( P_0 - \frac{\beta}{\varepsilon} ) \frac{a(a-1)}{2} \varepsilon^2
    ( 1- \varepsilon )^{a-2} \right ] \\~\\~\\
  &= P_3 ( 1- \varepsilon )^{a+1} + P_2 a\varepsilon ( 1- \varepsilon )^{a}
  + \frac{1}{2} P_1 {a ( a-1 ) } \varepsilon^2 ( 1- \varepsilon )^{a-1}
  + \frac{1}{6} ( P_0 - \frac{\beta}{\varepsilon} ) { ( a-1 ) [ ( a-1 )^2 - 1 ]}
  \varepsilon^3 ( 1- \varepsilon )^{a-2} \\
  &+ P_2 \varepsilon ( 1- \varepsilon )^a
  + P_1 a \varepsilon^2 ( 1- \varepsilon )^{a-1} + ( P_0 - \frac{\beta}{\varepsilon} )
  \frac{a(a-1)}{2} \varepsilon^3 ( 1- \varepsilon )^{a-2} \\~\\~\\
  &= P_3 ( 1- \varepsilon )^{a+1} + P_2 ( a+1 ) \varepsilon ( 1- \varepsilon )^{a}
  + \frac{1}{2} P_1 { ( a+1 ) a} \varepsilon^2 ( 1- \varepsilon )^{a-1}
  + ( P_0 - \frac{\beta}{\varepsilon} ) \varepsilon^3 ( 1- \varepsilon )^{a-2}
  \left [ \frac{ ( a-1 ) [ ( a-1 )^2 - 1]} {6} + \frac{a(a-1)}{2} \right ]
\end{aligned}
\]
\fontsize{12}{14}\selectfont
Simplifying the expression in the square brackets :
\fontsize{9}{12}\selectfont
\[
\begin{aligned}
  \frac{ ( a-1 )[ ( a-1 )^2 - 1]}{6} + \frac{a(a-1)}{2}
  &= \frac{1}{6} \left [ ( a-1 )[ ( a-1 )^2 - 1 ] + 3a( a-1 ) \right ]
  = \frac{1}{6} \left [ ( a-1 ) ( a^2 -2a + 1 - 1 + 3 a ) \right ] \\
  &= \frac{1}{6} \left [ ( a-1 ) ( a^2 + a ) \right ]
  = \frac{1}{6} \left [ a^3 - a^2 + a^2 - a \right ] = \frac{1}{6} ( a ( a^2 - 1 ) )
\end{aligned}
\]

\fontsize{12}{14}\selectfont
leads to : 
\fontsize{9}{12}\selectfont
\[
  {}^{a}{P_3} ( 1- \varepsilon ) + \varepsilon {}^{a}{ P_2} = P_3 ( 1- \varepsilon )^{a+1}
  + P_2 ( a+1 ) \varepsilon ( 1- \varepsilon )^{a}
  + \frac{1}{2} P_1 { ( a+1 ) a} \varepsilon^2 ( 1- \varepsilon )^{a-1}
  + \frac{1}{6} ( P_0 - \frac{\beta}{\varepsilon} ) \varepsilon^3 ( 1- \varepsilon )^{a-2}
  a( a^2 - 1) = {}^{a+1}{P_3}
\]

\fontsize{12}{14}\selectfont

And in a trivial way, when a=0 leads to ${}^0P_3 = P_3$. The formula  (\ref{P3d}) is then demonstrated.

\subsection{Proof of the convergence of $\sum_{n=2}^\infty \frac {n-1}{n!}$:}
\label{preuve_convergence}
\[
\begin{aligned}
  \sum_{n=2}^\infty \frac{n-1}{n!} &= \sum_{n=2}^\infty \frac{n}{n!} - \sum_{n=2}^\infty \frac{1}{n!} \\
  &= \sum_{n=2}^\infty \frac{1}{(n-1)!} - \sum_{n=2}^\infty \frac{1}{n!} \\
  &= \sum_{n=1}^\infty \frac{1}{n!}  - \sum_{n=2}^\infty \frac{1}{n!} \\
  &= \frac{1}{0!} + \sum_{n=2}^\infty \frac{1}{n!} - \sum_{n=2}^\infty \frac{1}{n!} = 1
\end{aligned}
\]

\subsection{Number of clusters after <<homogeneous nucleation>>:}
\label{preuve_nombre_homogene}
\[
\begin{aligned}
  \overline{n} = \sum_{n=2}^\infty s\Pi_n &= \sum_{n=2}^\infty \frac{n(n-1)}{n!} \\
  &= \sum_{n=2}^\infty \frac{(n-1)}{(n-1)!} \\
  &= \sum_{n=2}^\infty \frac{1}{(n-2)!} \\
  &= \sum_{n=0}^\infty  \frac{1}{n!} = e
\end{aligned}
\]

\medskip



\end{document}